\documentclass[conference]{IEEEtran}
\usepackage[utf8]{inputenc}
\usepackage{geometry}[margin=1in]
\usepackage{amsmath}
\usepackage{amssymb}
\usepackage{amsthm}
\usepackage{csvsimple}
\usepackage{graphicx}
\usepackage{hyperref}
\allowdisplaybreaks

\IEEEoverridecommandlockouts

\newtheorem{prop}{Proposition}

\title{A Nonlinear Sum of Squares
Search for CAZAC Sequences}
\author{Mark Magsino
\\ Department of Mathematics \\ U.S. Naval Academy \\ magsino@usna.edu

\and Yixin Xu \\ Department of Mathematics \\ Ohio
State University \\ xu.3520@buckeyemail.osu.edu}
\date{}

\DeclareMathOperator*{\argmin}{arg\,min}

\newcommand{\C}{\mathbb{C}}

\begin{document}

\maketitle

\begin{abstract}
We report on a search for CAZAC sequences by using 
nonlinear sum of squares optimization.
Up to equivalence, we found all length 7 CAZAC sequences.
We obtained evidence suggesting there are 
finitely many length 10 CAZAC
sequences with a total of 3040 sequences. Last, we compute
longer sequences and compare their aperiodic autocorrelation
properties to known sequences.
The code and results of this search are publicly
available through GitHub.
\end{abstract}

\section{Introduction}
Given $x \in \C^n$, we define the \textbf{periodic ambiguity
function} by
\begin{equation}
    A_p(x)[k,\ell] := \frac{1}{n} \sum_{j=0}^{n-1} x_{j+k}
    \overline{x_j}
    e^{-2 \pi i j \ell/n},
\end{equation}
where $0 \leq k,\ell \leq n-1$, and indices are taken modulo
$n$.
The \textbf{periodic autocorrelation}
is the case with no frequency shift, 
i.e. when $\ell = 0$. 
In this case it 
is convenient
to suppress the second input and write
\begin{equation}
A_p(x)[k] := A_p(x)[k,0] = \frac{1}{n}\sum_{j=0}^{n-1}x_{j+k}
\overline{x_j}.
\end{equation}
A
\textbf{CAZAC sequence of length $n$} as a vector $x \in \C^n$ with 
the properties
\begin{itemize}
	\item[(i)] $|x_j| = 1$ for $1 \leq j \leq n$,
	\item[(ii)] $ A_p(x)[k]= 0$ for $
	1 \leq k \leq n-1$.
\end{itemize}
The first property is known as the {\bf constant amplitude} property and
the second is known as the {\bf zero autocorrelation} property, 
which gives
rise to the acronym CAZAC. 

These sequences have several interpretations and applications which
motivate their study. In communication theory they have been used
to reduce the cross-correlation of signals, for 
uplink synchronization \cite{Mansour2009}, and OFDM for 
5G communication \cite{Feng2015}\cite{Wesolowski2015}. It is also
studied as an idealized waveform with regards to the narrowband
ambiguity function in radar \cite{Vakman1968}. In general,
the constant
amplitude allows one to encode information purely in terms of
phase and the zero autocorrelation ensures no interference
with shifted copies of the signal. CAZAC sequences
are also known as \textbf{perfect polyphase sequences}
and are also well studied under that name
\cite{Mow1995}\cite{Mow1992}\cite{Park2015}.

CAZAC sequences have perfect periodic autocorrelation, 
but many applications require their
aperiodic autocorrelation.
The \textbf{aperiodic ambiguity function} of
a sequence $x \in \C^n$ is given by
\begin{equation}
    A_a(x)[k,\ell] := \sum_{j=1}^{n-1} 
    x^{(a)}_{j+k}\overline{x^{(a)}_{j}}
    e^{-2 \pi i j \ell / n},
\end{equation}
where $0 \leq k,\ell \leq n-1$ and $x^{(a)}_j$ is
defined by
\begin{equation}
    x_j^{(a)} = \begin{cases}
    x_j, & \text{ if } 0 \leq j \leq n-1, \\
    0, & \text{ otherwise}.
    \end{cases}
\end{equation}
The key difference is instead of taking indices modulo $n$,
we set the value to zero when the index falls outside of
the usual range. The \textbf{aperiodic autocorrelation}
corresponds to $\ell = 0$ and it is convenient
to write
\begin{equation}
    A_a(x)[k]:= A_a[k,0]
    =\sum_{j=1}^{n-1} 
    x^{(a)}_{j+k}\overline{x^{(a)}_{j}}.
\end{equation}
In particular, we use aperiodic autocorrelation to study two 
properties of interest. We define the \textbf{peak sidelobe
level (PSL)} of $x \in \C^n$ by
\begin{equation}
    \mathrm{PSL}(x) = \frac{1}{|A_a(x)[k]|}\max_{k \neq 0} |A_a[k]|,
\end{equation}
and the \textbf{integrated sidelobe level (ISL)} by
\begin{equation}
    \mathrm{ISL}(x) = \frac{1}{|A_a(x)[k]|^2}
    \sum_{k=1}^{n-1} |A_a[k]|^2.
\end{equation}

Although CAZAC sequences are well studied, many
things about them are still unknown. 
Given  two CAZAC sequences, $x$ and $y$, we say they are
equivalent if there exists a complex scalar $c$ with $|c|=1$
so that $y = cx$ and make the representative
of the equivalence class the sequence whose first entry
is 1. With this in mind, it is natural to ask: 
For each $n$, how many
CAZAC sequences of length $n$ are there?
As a partial answer, it is known that if $n$ is
prime, then there are at most $\binom{2n-2}{n-1}$ 
CAZAC sequences \cite{Haagerup2008}. 
If $n$ is composite and divisible by a perfect square,
then there are infinitely many sequences 
\cite{Bjorck1995}\cite{Milewski1983}.
If $n$ is composite and not divisible by any perfect square,
then it is unknown how many there are. 
A brute force calculation verifies
that there are finitely many for $n=6$
\cite{Bjorck1991}. Beyond that, it is
currently unknown.

There are additional transformations under which
CAZAC sequences are closed \cite{Benedetto2007}. They are a 
finite set of transformations so it does not
fundamentally change the question of whether the set of CAZAC
sequences of a given length is finite. We use these
to filter out known length 7 CAZAC sequences.

\begin{prop}\label{cazacclose}
Let $x \in \C^n$ be a CAZAC sequence and let
$\omega = e^{2 \pi i / n}$. Then, the following sequences
are also CAZAC sequences:
\begin{itemize}
    \item[(i)] $(T_kx)_j = x_{j+k}$, $0 \leq k \leq n-1$,
    \item[(ii)] $(M_\ell x)_j = \omega^{\ell j}x_j$,
    $0 \leq \ell \leq n-1$,
    \item[(iii)] $(D_m x)_j = x_{mj}, \mathrm{gcd}(m,n) = 1,$
    \item[(iv)] $(\overline{x})_j = \overline{x_j}$.
\end{itemize}
\end{prop}
\section{CAZAC Sequneces of Length 7}
The CAZAC sequences of length 7
can be split into
quadratic phase sequences and non-quadratic
phase sequences. Suppose $x\in \C^n$ is defined by
\[
x_j = e^{\pi i p(j)/n},
\]
where $p(j)$ is a quadratic polynomial. In this case,
we say that $x$ is a \textbf{quadratic
phase sequence}. The polynomials associated with the
known quadratic phase CAZAC sequences are
\begin{align*}
    \textit{Zadoff-Chu: } & p(j) = j(j-1), \hspace{0.5em} 
    (n \text{ odd})
    \\
    \textit{P4: } & p(j) = j(j-n).\\
    \textit{Wiener: } & p(j) = 2kj^2, \hspace{0.5em}
    (\mathrm{gcd}(k,n) = 1, n \text{ odd}),\\
    ~ & p(j) = kj^2, \hspace{0.5em}
    (\mathrm{gcd}(k,2n)=1, n \text { even}).
\end{align*}
When $n$ is prime,
there are at least $n(n-1)$ CAZAC sequences comprised of roots
of unity, including the quadratic phase sequences \cite{Benedetto2019}.
When $n=7$, this gives at least 42 roots of unity sequences. 
Moreover, the transformations described in Proposition \ref{cazacclose}
will keep the sequence a root of unity sequence.
On the other hand, in 
\cite{Bjorck1990}
Bj\"{o}rck constructed CAZAC sequences comprised of non roots of
unity for each prime $p > 5$. The construction is as follows.

Given an odd prime $p$,
let $\left(\frac{j}{p}\right)$ denote the \textbf{Legendre symbol}
defined by
\[
\left(\frac{j}{p}\right) = 
\begin{cases}
0, & \!\!\!\!\text{ if } j \equiv 0 \mod p, \\
1, & \!\!\!\!\text{ if } j \equiv x^2 \!\!\!\!\!
\mod p, \text{ for some } x \neq 0, \\
-1, & \!\!\!\!\text{ if } j \not\equiv x^2 \mod p, \text{ for any } x 
\neq 0.
\end{cases}
\]
We define the \textbf{Bj\"{o}rck sequence} of length $p$ by
\begin{equation}
x_j = e^{i \theta(j)}, \hspace{0.5em} 0 \leq j \leq p-1,
\end{equation}
where if $p \equiv 1 \mod 4$, then $\theta(j)$ is given by
\begin{equation}
\left(\frac{j}{p}\right)\mathrm{arccos}\left(
\frac{1}{1+\sqrt{p}}
\right),
\end{equation}
and if $p \equiv 3 \mod 4$, then $\theta(j)$ is given by
\begin{equation}
\theta(j) = \begin{cases}
\mathrm{arccos}\left(\frac{1-p}{1+p}\right) & \text{ if }
\left(\frac{j}{p}\right) = -1, \\
0, & \text{ otherwise}.
\end{cases}
\end{equation}
Since $7 \equiv 3 \mod 4$, the Bj\"{o}rck sequence of length 7 is
\begin{equation}
x = (1,1,1,e^{i\theta_7},1,e^{i\theta_7},e^{i\theta_7}),
\end{equation}
where $\theta_7 = \mathrm{arccos}(3/4)$.
Since CAZAC sequences are closed
under the operations outlined in Proposition \ref{cazacclose} 
the Bj\"{o}rck sequence
generates up to 252 CAZAC sequences which begin with 1 when p = 7. 
Some combinations of transformations result in the same sequence
and thus this is an over count, which our computations reflect.
Regardless, a total of at most 294 of the
532 sequences are accounted for, leaving many new sequences
of length 7 to be found.

\section{CAZACs as an Algebraic Variety}
The conditions which define a CAZAC sequence can be viewed
as a system of $2n-1$ equations with $n$ variables. 
A natural
question is to ask about the dimension of the set of 
solutions to this system. If
the set of solutions is zero dimensional, then it is a
discrete set and may be a finite set.
Conversely, if the set of solutions has positive dimension,
then there are infinitely many of
them. This question puts the problem in the realm of algebraic geometry.

The techniques of algebraic geometry only work with
systems of polynomials. However, the conjugations in the
sums prevent those conditions from being interpreted as polynomials.
We can get around this by expressing entries in
real and imaginary parts and converting the equations to polynomials of
real variables. Given a vector $X \in \C^n$
and expressing its entries as $x_j = a_j + i b_j$, CAZAC sequences
arise as solutions to the system
\begin{align}
    a_j^2 + b_j^2 = 1, & \hspace{1em} 0 \leq j \leq n-1, 
    \label{cazacsys1}\\ 
    \sum_{j=0}^{n-1} a_{j+k}a_j + b_{j+k}b_j = 0, 
    & \hspace{1em} 1 \leq k \leq n-1, \\
    \sum_{j=0}^{n-1} a_j b_{j+k} - b_j a_{j+k} = 0,
    & \hspace{1em} 1 \leq k \leq n-1. \label{cazacsys2}
\end{align}

Since constraining each amplitude has $n$ conditions, and
each of the $n-1$ autocorrelation conditions has been split into
real and imaginary parts, there are $3n-2$ equations in this system.
Since each entry of the sequence was split into a real and
imaginary part, the system has $2n$ variables. Hence,
the system is a real algebraic variety with $3n-2$ equations and $2n$
variables.
\section{Nonlinear Least Squares Search}
The system of equations
(\ref{cazacsys1}) - (\ref{cazacsys2}) can be converted into an
unconstrained nonlinear sum of squares optimization problem. 
Let $a = (a_1,\cdots,a_n)$ and $b = (b_1,\cdots,b_n)$, and
for 
each $1 \leq k
\leq n-1$ and $0 \leq \ell \leq n-1$, define the functions
\begin{align}
    f_\ell(a,b) & = a_\ell^2 + b_\ell^2 - 1, \\
    g_k(a,b) &= \sum_{j=0}^{n-1} a_{j+k}a_j + b_{j+k}b_j, \\
    h_k(a,b) &= \sum_{j=0}^{n-1} a_j b_{j+k} - b_j a_{j+k}.
\end{align}
Note that $(a,b)$ are the real and imaginary parts, respectively,
of a CAZAC sequence precisely when all $3n-2$ functions are all
zero at $(a,b)$. We can obtain CAZAC sequences as
solutions to the unconstrained optimization problem
\begin{equation}
    \argmin_{(a,b)} \sum_{\ell=0}^{n-1}f_\ell^2
    + \sum_{k=1}^{n-1} g_k^2 + h_k^2, \label{argminchar}
\end{equation}
for which the objective function is zero. 

Since the equations are nonlinear polynomials, optimization
procedures are not guaranteed to work. Specifically, there
may be local minima where
the objective function is greater than zero. Moreover,
the objective function is not convex which makes it
difficult to give theoretical guarantees
about this method.
Nonetheless, in our experiments we found very few such local minima
and generated as many sequences as needed for each test.

\section{Method}
The experimental design was implemented in a Jupyter Notebook.
All trials were run on a 2020 MacBook Pro with the 10-core M1 Pro CPU
and 16 GB of physical memory. The optimization problem in
(\ref{argminchar}) was implemented using 
the least squares solver in SciPy's
Optimization package. By default the solver uses the
Trust Region Reflective (TRR) algorithm and a finite difference
operation to estimate the Jacobian at each point. The code is
available through GitHub at:
\href{https://github.com/magsino-usna/IEEE-SoS-CAZAC.git}
{https://github.com/magsino-usna/IEEE-SoS-CAZAC.git}.

\subsection{Length 7 CAZACS}
We perform the optimization on 10,000 random initial starting
points in $[-1,1]^{14}$. For
increased precision, we used stronger tolerances of $10^{-12}$ instead
of the default $10^{-8}$ for changes in the gradient, input
variables, and cost function. If the resulting cost function
was lower than $10^{-10}$, we considered the point found as a 
CAZAC sequence stored it as a new row in an array.

After generating the list of sequences, we rounded each one to
8 decimal places and pass the rounded list through 
NumPy's algorithm
for determining unique rows in arrays to determine the number of
sequences found.
After that, we used the transformations listed in Proposition
\ref{cazacclose} using the Bj\"{o}rck and the Wiener sequences as
base sequences to filter out all previously known CAZAC sequences.

\subsection{Length 10 CAZACs}\label{l10_method}
Since 10 is composite and not divisible by a perfect square, 
it is not known if there are finitely many of them. 
Our idea for exploring whether the set of solutions 
is finite is
based on the following intuition. 

\begin{figure}[ht]
\centering
\includegraphics[width=0.49\textwidth]{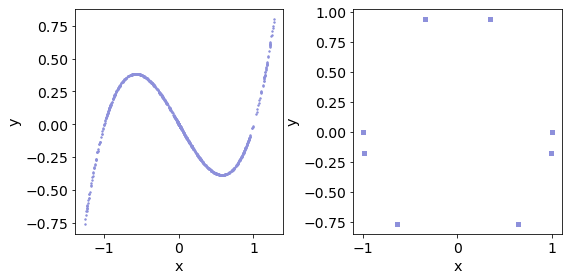}
\caption{Experiments taking 1000 initial points and using
nonlinear sum-of-squares to find points onto two sets. Left:
The points traced out the one dimensional curve
$x^3-x-y=0$. 
Right: The points went to the 8 points of the
intersection of $x^2+y^2-1=0$ and $8x^4-10x^2-x+2=0$.}
\label{projection_figure}
\end{figure}

Solving the sum of squares minimization with an initial point is
essentially projecting the point onto the
set of solutions. If the set of solutions is greater than zero
dimensional, most of
the initialization points will project to 
different points
on the continuous set of solutions. Conversely, if the set
of solutions is zero dimensional and finite, then most of the points
should project to that finite set of points. Thus, if the number
of unique sequences found is far smaller than the number of trials,
the set is likely finite.
This idea is illustrated
in Fig. \ref{projection_figure}.

\vspace{-1em}
For the experiment, we performed the optimization on 200,000
random initial starting points in $[-1,1]^{20}$ and use the
same tolerances and rounding method as in the length 7 case.

\subsection{Aperiodic Autocorrelations
of Longer CAZACs}

Using the nonlinear sum of squares optimization method, we computed
1000 CAZAC sequences of prime lengths $n = 11, 13, 17, 23, 29, 37, 43,
47$. We then compared the PSL and ISL of the CAZAC sequences to
the PSL and ISL of the corresponding Zadoff-Chu and Bj\"{o}rck 
sequences. We compare the results using a box plot to give a sense
of how generic CAZAC sequences behave with respect to PSL and ISL
behavior.

\section{Results}
As previously mentioned, all results and code
are stored on GitHub:
The full list is available on GitHub:
\href{https://github.com/magsino-usna/IEEE-SoS-CAZAC.git}
{https://github.com/magsino-usna/IEEE-SoS-CAZAC.git}.
\subsection{Length 7 Sequences}
\begin{figure}[ht]
\centering
\includegraphics[width=0.48\textwidth]{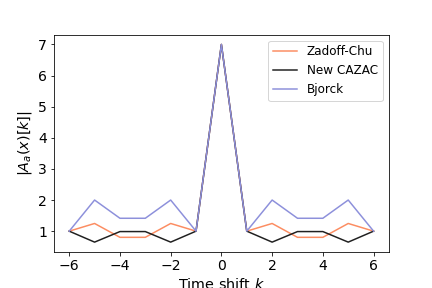}
\caption{Autocorrelation of the best performing length 7 CAZAC
sequence plotted against the autocorrelations of the length 7
Zadoff-Chu and Bj\"{o}rck sequences. The new CAZAC sequence achieves
a lower PSL than the Zadoff-Chu sequence.}
\label{aac7}
\end{figure}

We were able to enumerate all 532 CAZAC, handling equivalence by
dividing and making the first entry 1. This took 15 min
to run all 100,000 attempts on the aforementioned MacBook Pro. 
The maximum final value of the objective function across all 
trials was on the order of $10^{-17}$ which
implies the objective function has no spurious local minima in
the length 7 case. We picked out the new CAZAC with the best
PSL and compared its autocorrelation to the Zadoff-Chu and
Bj\"{o}rck sequences in Fig. \ref{aac7}.

\subsection{Length 10 Sequences}
We ran three sets of the trials described in Section \ref{l10_method}.
In all three trials,
the same 3040 CAZAC sequences were found. This heavily implies
that there are 3040 CAZAC sequences of length 10, although it is
still possible that points corresponding to a highly unstable minimum
could be missed. However, the average runtime for each set of these trials
was about 13 hours. The maximum cost
function value across all trials was to the order of $10^{-24}$, which
implies once again that every local minimum of the objective function
is the global minimum of 0. All 3040 sequences found are also
available through GitHub as well.

\subsection{Aperodic Correlations}
\begin{figure}[ht]
\centering
\includegraphics[width=0.48\textwidth]{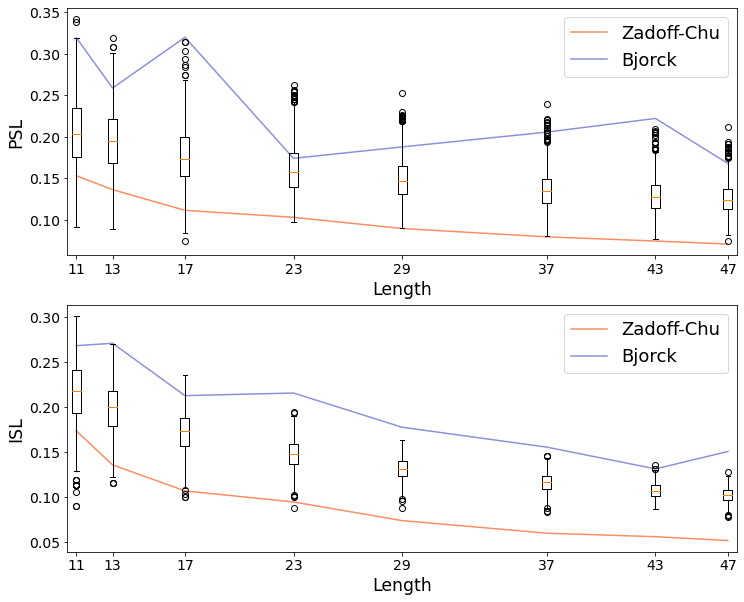}
\caption{
Top: The PSLs of each numerically found CAZAC 
are compared to the Zadoff-Chu and
Bj\"{o}rck sequences by box plot.
Bottom: The ISLs of each numerically found CAZAC are 
compared to the Zadoff-Chu and
Bj\"{o}rck sequences by box plot.
Notably, it seems possible to get better PSL and ISL
than the Zadoff-Chu sequence in some cases.}
\label{psl_isl}
\end{figure}

We were successfully able to find 1,000 CAZAC sequences of
each length $n$ = 11,13,17,23,29,37,43,47. For these larger
cases a few spurious local minima were found in initial tests requiring
more minimizations than the number of desired sequences,
slowing down the computation. We
compare the PSL and ISL properties of the numerical
CAZAC sequences found
with the box plot in
Fig. \ref{psl_isl}.

\begin{figure}[ht]
\centering
\includegraphics[width=0.40\textwidth]{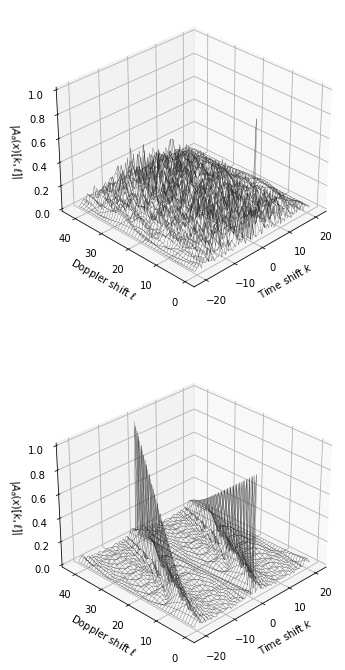}
\caption{Top: Ambiguity function plot for a computed length 43 
CAZAC sequence.
Bottom: Ambiguity function plot for the length 43 Zadoff-Chu 
sequence. Although not as flat
as the Zadoff-Chu sequence, the energy is more well spread in the
computed sequence so no values exceed 0.3 times the ambiguity function
at $(k,\ell)=(0,0)$.}
\label{43af}
\end{figure}

To further illustrate the properties of 
CAZAC sequences generated
by these methods, we juxtapose the (modulus) of 
the ambiguity function of a length 43 numerically found 
CAZAC sequence and the typically used
used Zadoff-Chu sequence. This is depicted in Fig. \ref{43af}.

\section{Future Directions}
This work provides a framework for computing CAZAC sequences by
using nonlinear sum of squares optimization. In addition to
the potential problem of spurious local minima, scalability
is an issue as well. It is possible that other optimization methods
or equation solvers would lead to improved efficiency.
The
code has been made available through GitHub to help
ensure replicability of the numerical experiments and to provide
a base of code for future directions of work.

It is also likely that the system of equations defining CAZAC
sequences has some symmetries or redundancies. Exploring these could
make it possible to reduce the number of equations and find
new sequences more efficiently.

Another alternative is to convert the aperiodic autocorrelations
into a system of polynomials of real variables. We could then use the
resulting function to create objective functions that act as a
proxy for PSL or ISL and run an optimization search. This could
also give us insight on what the optimal PSL and ISL values are
for phase-coded waveforms.

Since this work involves an algebraic variety, it is useful
to consider techniques from numerical algebraic geometry. For example,
homotopy continuation \cite{Alexander1978} is a framework
for analyzing isolated solutions of algebraic varieties. This could
be useful in determining whether there are finitely many CAZAC
sequences in currently unknown cases.
\section*{Acknowledgements}
The views expressed 
in the paper are those of the first author and
do not reflect the official policy or position of the Department
of the Navy, Department of Defense, or the U.S. Government.

\bibliographystyle{plain}
\bibliography{numeric_cazac}

\end{document}